\documentclass[pra,10pt,twocolumn,superscriptaddress,showpacs,floatfix]{revtex4-1}
\usepackage{amsmath}
\usepackage{latexsym}
\usepackage{amssymb}
\usepackage{graphicx}
\usepackage[colorlinks=true, citecolor=blue, urlcolor=blue]{hyperref}
\usepackage{float}
\usepackage{graphics}
\usepackage{amsfonts}

\begin{document}
\title{Device independent Schmidt rank witness by using Hardy paradox}

\author{Amit Mukherjee}
\affiliation{Physics and Applied Mathematics Unit, Indian Statistical Institute, 203 B. T. Road, Kolkata 700108, India.}

\author{Arup Roy}
\affiliation{Physics and Applied Mathematics Unit, Indian Statistical Institute, 203 B. T. Road, Kolkata 700108, India.}

\author{Some Sankar Bhattacharya}
\affiliation{Physics and Applied Mathematics Unit, Indian Statistical Institute, 203 B. T. Road, Kolkata 700108, India.}

\author{Subhadipa Das}
\affiliation{S.N. Bose National Center for Basic Sciences, Block JD, Sector III, Salt Lake, Kolkata 700098, India.}

\author{Md. Rajjak Gazi}
\affiliation{Murshidabad Institute of Technology, Cossimbazar, Murshidabad 742102, India.}

\author{Manik Banik}
\email{manik11ju@gmail.com}
\affiliation{Physics and Applied Mathematics Unit, Indian Statistical Institute, 203 B. T. Road, Kolkata 700108, India.}

\begin{abstract}
Schmidt rank of bipartite pure state serves as a testimony of entanglement. It is a monotone under local operation $+$ classical communications (LOCC) and puts restrictions in LOCC convertibility of quantum states. Identifying the Schmidt rank of an unknown quantum state therefore seek importance from information theoretic perspective. In this work it is shown that a modified version of Hardy's argument, which reveals the contradiction of quantum theory with \emph{local realism}, turns out to be useful for inspecting the minimal Schmidt rank of the unknown state and hence also the minimal dimension of the system. Use of Hardy's test in such task provides a practical advantage: the Schmidt rank can be determined without knowing the detailed functioning of the experimental devices i.e., Hardy's test suffices to be a device independent Schmidt rank witness.      
\end{abstract}


\maketitle
Among various counterintuitive features of quantum mechanics, certainly, one of the most bizarre property is quantum entanglement \cite{Schrodinger,Horodecki,Werner}. This holistic property of compound quantum systems, which involves non-classical correlations among subsystems, has potential for many quantum processes, including canonical ones: quantum cryptography \cite{Crypto}, quantum teleportation \cite{Tele}, and dense coding \cite{Dense}. According to the quantum formalism, the total Hilbert space $\mathcal{H}$ of the $n$ separate systems is a tensor product of the subsystem spaces, i.e., $\mathcal{H}=\otimes_{k=1}^n\mathcal{H}_k$. When the number of the involved subsystems are two, the pure state $|\psi\rangle$ of the bipartite system can always be described by its Schmidt decomposition, i.e., the representation of $|\psi\rangle$ in an orthogonal product basis with minimal number of terms \cite{Peres}. A bipartite pure state $|\psi\rangle\in\mathcal{H}_1\otimes\mathcal{H}_2$, with $\mbox{dim}\mathcal{H}_1=d_1$ and $\mbox{dim}\mathcal{H}_2=d_2\ge d_1$, has Schmidt rank $s$ if its Schmidt decomposition reads: $|\psi\rangle=\sum_{i=1}^{s}\alpha_i|e_i\rangle\otimes|f_i\rangle$, where $s\le d_1$, $\sum_{i=1}^{s}\alpha^2_i=1$, $\alpha_i>0$ and $\{|e_i\rangle\}_{i=1}^{d_1}\subset\mathcal{H}_1$ is an orthonormal set of vectors in the Hilbert space $\mathcal{H}_1$ and $\{|f_i\rangle\}_{i=1}^{d_1}\subset\mathcal{H}_2$ is an orthonormal set of vectors in the Hilbert space $\mathcal{H}_2$. The Schmidt number for a bipartite mixed state $\rho\in\mathcal{D}(\mathcal{H}_1\otimes\mathcal{H}_2)$ is the number $k$ such that: (a) for any decomposition $\{p_i\ge0,|\psi_i\rangle\}$ of $\rho$ with $\rho=\sum_ip_i|\psi_i\rangle\langle\psi_i|$ at least one of the vectors $\{|\psi_i\rangle\}$ has at least Schmidt rank $k$, and (b) and  there exists a decomposition of $\rho$ with all vectors $\{|\psi_i\rangle\}$ of Schmidt rank at most $k$ \cite{Terhal}. Here, $\mathcal{D}(\mathcal{H}_1\otimes\mathcal{H}_2)$ denotes the collection of positive, trace-$1$ operators acting on $\mathcal{H}_1\otimes\mathcal{H}_2$. 

The Schmidt rank is the number of non vanishing terms in Schmidt decomposition. This decomposition gives a clear insight into the number of degrees of freedom that are entangled between both parties --- if the Schmidt rank is greater than unity then the pure bipartite state must be entangled. Furthermore it has been proved that the Schmidt number is non-increasing under local operations and classical communication (LOCC), i.e., it is a monotone under LOCC. Hence, it puts restriction in LOCC convertibility of states \cite{Nielsen0,Plenio}. A necessary condition for a pure state to be convertible by LOCC to another pure state is that the Schmidt rank of the later cannot be larger than that of the previous one \cite{Lo}.

From an information theoretic point of view, the Schmidt rank of a state, therefore, can be considered as a resource. Identifying the Schmidt rank of a state is also important for quantifying the power of quantum correlations, a central issue in Quantum Information Theory. In this work the problem of determining the Schmidt rank of an unknown bipartite state has been addressed. Interestingly, it has been shown that considering a modified version of the Hardy's paradox, recently introduced by Chen \emph{et.al} \cite{Chen}, one can know the minimal Schmidt rank of the given unknown state. It also provides information about the minimal Hilbert space dimension of the concerned system. The use of nonlocality argument in this task comes up with a novel advantage. The Schmidt rank can be determined from measurement data alone, in a scenario in which all devices used in the experiment, including the measurement device, are uncharacterized or in other words no assumption about the internal working of the devices is needed. 

The original Hardy's argument was defined for \emph{dichotomic} observables, i.e., observables with two outcomes \cite{Hardy1,Hardy2}. The authors in Ref.\cite{Chen} have generalized it for observables with arbitrary many outcomes. Moreover, they have shown that, unlike the original Hardy's argument, the success probability of the many-outcome argument increases with the increase of the system's dimension. In this work the three-outcome Hardy's argument has been considered. Firstly, it has been shown that neither a $2$-qubit state nor a qubit-qutrit state exhibits this argument for three-outcome generalized measurement, i.e., positive-operator-valued-measurement (POVM) \cite{Kraus,Nielsen}. Using this result it has been further shown that this argument can be designed as a device independent Schmidt rank as well as Hilbert space dimension witness. 

Before going to the main result a quick overview on the Hardy's argument has been presented. L. Hardy provided an elegant argument which, like Bell's inequality \cite{Bell}, reveals nonlocality within quantum mechanics \cite{Hardy1,Hardy2} and it is commonly called `Hardy paradox'. Hardy's proof is usually considered ``the simplest form of Bell’s theorem'' \cite{Mermin}. The argument requires two spatially separated observers, say Alice and Bob, each with two measurements (the measurements for Alice and Bob are denoted by $A_i$ and $B_j$ respectively, with $i,j\in\{1,2\}$), each with two possible outcomes denoted by `$0$' and `$1$'. It puts restrictions on a certain choice of $4$ out of $16$ joint probabilities in the correlation matrix. One such choice is:
\begin{subequations}
\begin{align}\label{H1}
P(A_1=1,B_2=1) = 0,\\\label{H2}
P(A_2=1,B_1=1) = 0,\\\label{H3}
P(A_1=0,B_1=0) = 0,\\ \label{H4}
P(A_2=1,B_2=1) > 0.
\end{align}\label{Hardy}
\end{subequations}
Here $P(A_i=m,B_j=n)$ denotes the conditional joint probability of obtaining outcome `$m$' by Alice and outcome `$n$' by Bob when they perform measurement $A_i$ and $B_j$, respectively; $m,n\in\{0,1\}$. The non zero probability in Eq.(\ref{Hardy}) (i.e. left hand side of Eq.(\ref{H4})) is called Hardy's success probability, $P_{Hardy}=P(A_2=1,B_2=1)$. For $2$-qubit system the maximum achievable value of Hardy's success is $P_{Hardy}=\frac{5\sqrt{5}-11}{2}\approx0.09$ \cite{Hardy2}. It is important to note that for $2$-qudit system this maximum success probability remains same \cite{Seshadreesan,Rabelo}, i.e., for showing the contradiction of quantum mechanics with \emph{local realism}, higher dimensional systems give no advantage in experimental implementation of such a test. 

Recently, the authors in \cite{Chen} have introduced a Hardy like argument for $d$-outcome measurements, i.e., $m,n\in\{0,1,....,d-1\}$. Denoting the joint conditional probability as  $P(A_2<B_1)=\sum_{m<n}P(A_2=m,B_1=n)$ the argument reads as $P(A_2<B_1)=0$, $P(B_1<A_1)=0$, $P(A_1<B_2)=0$, $P(A_2<B_2)>0$. For two outcomes the argument boils down to the original Hardy's argument, i.e. Eq.(\ref{Hardy}). For three outcomes their argument explicitly looks:
\begin{subequations}
\begin{align}\label{H31}
P(A_2=0,B_1=1)+P(A_2=0,B_1=2)\nonumber\\
+P(A_2=1,B_1=2)=0\\\label{H32}
P(A_1=1,B_1=0)+P(A_1=2,B_1=0)\nonumber\\
+P(A_1=2,B_1=1)=0\\\label{H33}
P(A_1=0,B_2=1)+P(A_1=0,B_2=2)\nonumber\\
+P(A_1=1,B_2=2)=0\\\label{H34}
P(A_2=0,B_2=1)+P(A_2=0,B_2=2)\nonumber\\
+P(A_2=1,B_2=2)>0
\end{align}\label{Hardy3}
\end{subequations}
Likewise (\ref{H4}), the left hand side in (\ref{H34}) measures the success probability of three-outcome Hardy's test and similarly for the $d$-outcome cases. Higher value of this quantity implies that experimentally it is easier to demonstrate the contradiction of quantum mechanics with \emph{local realism}. 

In quantum theory the joint conditional probabilities are calculated as:
\begin{equation*}
P(A_i=m,B_j=n)=\mbox{Tr}(\mathcal{E}^m_{A_i}\otimes \mathcal{F}^n_{B_j}\rho_{AB}),
\end{equation*}
where, $\{\mathcal{E}^m_{A_i}~|~\mathcal{E}^m_{A_i}>0~\forall~m,~\sum_m\mathcal{E}^m_{A_i}=\mathbf{1}_{\mathcal{H}_A}\}$ and $\{\mathcal{F}^n_{B_j}~|~\mathcal{F}^n_{B_j}>0~\forall~n,~\sum_n\mathcal{F}^n_{B_j}=\mathbf{1}_{\mathcal{H}_B}\}$ are POVMs acting on Alice's and Bob's side respectively and $\rho_{AB}\in\mathcal{D}(\mathcal{H}_A\otimes\mathcal{H}_B)$ is the shared state between Alice and Bob. Considering $2$-qudit pure states and projective measurements the authors in \cite{Chen} find the optimal success probability for $d$-outcome Hardy's test. As for example, for $3$-outcome case the optimal achievable success probability is $0.141327$, which is strictly greater than the optimal two-outcome Hardy's success probability. This value can be achieved by performing three-outcome projective measurements on $2$-qutrit system. It seems to imply that the success probability increases with increasing system's dimension. However this implication is not conclusive. Cause it has not yet been proved that by sharing a $\mathbb{C}^2\otimes\mathbb{C}^2$ state (or $\mathbb{C}^2\otimes\mathbb{C}^3$ state) and performing three-outcome generalized measurement one cannot exhibit the Hardy's paradox (\ref{Hardy3}) with success probability greater than $0.09$. In the following a more powerful result has been proved -- that neither a $2$-qubit state nor a qubit-qutrit state exhibits the three-outcome Hardy's paradox (\ref{Hardy3}).  

First the 2-qubit case has been  considered. Let Alice and Bob share a 2-qubit state. Both of them perform two three-outcome POVMs, with outcomes denoted by $m,n\in\{0,1,2\}$ respectively. Let $\mathcal{E}^m_i$ denote the POVM element corresponding to Alice's outcome $m$ and similarly $\mathcal{F}^n_j$ for Bob's outcome $n$. Thus we have:
\begin{equation*}
\sum_{m=0}^2\mathcal{E}^m_i=\mathbf{1}_2,~~\sum_{n=0}^2\mathcal{F}^n_j=\mathbf{1}_2;~~\mbox{for}~~i,j\in\{1,2\}.
\end{equation*} 
There are following possible cases: (i) All the POVM elements are rank one operators; (ii) Some of the POVM elements may have more than one rank. In the course of analysis, intuitively, it will become clear that if the measurements with rank one POVM elements do not pass the three-outcome Hardy's test then it is even more difficult for the measurements with higher rank POVM elements to pass it.

{\bf Case (i)}: In this case all the POVM elements can be considered as proportional to projection operators on some ray vectors, i.e.,
\begin{equation*}
\mathcal{E}^m_i\propto\Pi[\psi_i^m],~~\mbox{and}~~\mathcal{F}^n_j\propto\Pi[\phi_j^n],
\end{equation*}  
where $\Pi[\psi_i^m]\equiv|\psi_i^m\rangle\langle\psi_i^m|$ and $\Pi[\phi_j^n]\equiv|\phi_j^n\rangle\langle\phi_j^n|$. The 2-qubit state that exhibits the Hardy's argument (\ref{Hardy3}) must satisfy the conditions (\ref{H31})-(\ref{H33}), which imply that each term on the left hand side of these equations must be zero. The condition $P(A_2=0,B_1=1)=\mbox{Tr}(\mathcal{E}^0_{2}\otimes \mathcal{F}^1_{1}\sigma_{AB})=0$ implies that the concerned 2-qubit state is orthogonal to the product vector $|\psi_2^0\rangle\otimes|\phi_1^1\rangle$ and similar is true for other cases. Thus the conditions (\ref{H31})-(\ref{H33}) altogether imply that the concerned state must be orthogonal to the following nine vectors:
$$\mathcal{\bf S}\equiv\{\mathcal{V}_1=|\psi_2^0\rangle\otimes|\phi_1^1\rangle,\mathcal{V}_2=|\psi_2^0\rangle\otimes|\phi_1^2\rangle,$$
\vspace{-.8cm}
$$\mathcal{V}_3=|\psi_2^1\rangle\otimes|\phi_1^2\rangle,\mathcal{V}_4=|\psi_1^1\rangle\otimes|\phi_1^0\rangle,$$
\vspace{-.8cm}
$$\mathcal{V}_5=|\psi_1^2\rangle\otimes|\phi_1^0\rangle, \mathcal{V}_6=|\psi_1^2\rangle\otimes|\phi_1^1\rangle,$$
\vspace{-.8cm}
$$\mathcal{V}_7=|\psi_1^0\rangle\otimes|\phi_2^1\rangle,\mathcal{V}_8=|\psi_1^0\rangle\otimes|\phi_2^2\rangle, $$
\vspace{-.8cm}
$$\mathcal{V}_9=|\psi_1^1\rangle\otimes|\phi_2^2\rangle\}$$
Without loss of generality, consider $\{|\psi_1^0\rangle,|\psi_1^1\rangle\}$ as the basis for Alice's qubit system. Other states on Alice's side when written in linear combination of this basis, read as:
\begin{subequations}
\begin{align}\label{A1}
|\psi_1^2\rangle=\alpha_1^2|\psi_1^0\rangle+\beta_1^2|\psi_1^1\rangle,\\\label{A2}
|\psi_2^m\rangle=\alpha_2^m|\psi_1^0\rangle+\beta_2^m|\psi_1^1\rangle. 
\end{align}\label{A}
\end{subequations}
Similarly choosing $\{|\phi_1^0\rangle,|\phi_1^1\rangle\}$ as the basis for Bob's qubit:
\begin{subequations}
\begin{align}\label{B1}
|\phi_1^2\rangle=\delta_1^2|\phi_1^0\rangle+\gamma_1^2|\phi_1^1\rangle,\\\label{B2}
|\phi_2^n\rangle=\delta_2^n|\phi_1^0\rangle+\gamma_2^n|\phi_1^1\rangle. 
\end{align}\label{B}
\end{subequations}
Consider the set $\{|\psi_1^0\rangle\otimes|\phi_1^0\rangle,|\psi_1^0\rangle\otimes|\phi_1^1\rangle,|\psi_1^1\rangle\otimes|\phi_1^0\rangle,
|\psi_1^1\rangle\otimes|\phi_1^1\rangle\}$ as the basis for the tensor product space $\mathbb{C}^2\otimes\mathbb{C}^2$. The nine vectors in the set $\mathcal{S}$ written in the above basis read as:
\begin{subequations}
\begin{align}\label{vec1}
\mathcal{V}_1\equiv(0,\alpha_2^0,0,\beta_2^0),\\\label{vec2}
\mathcal{V}_2\equiv(\alpha_2^0\delta_1^2,\alpha_2^0\gamma_1^2,\beta_2^0\delta_1^2,\beta_2^0\gamma_1^2),\\\label{vec3}
\mathcal{V}_3\equiv(\alpha_2^1\delta_1^2,\alpha_2^1\gamma_1^2,\beta_2^1\delta_1^2,\beta_2^1\gamma_1^2),\\\label{vec4}
\mathcal{V}_4\equiv(0,0,1,0),\\\label{vec5}
\mathcal{V}_5\equiv(\alpha_1^2,0,\beta_1^2,0),\\\label{vec6}
\mathcal{V}_6\equiv(0,\alpha_1^2,0,\beta_1^2),\\\label{vec7}
\mathcal{V}_7\equiv(\delta_2^1,\gamma_2^1,0,0),\\\label{vec8}
\mathcal{V}_8\equiv(\delta_2^2,\gamma_2^2,0,0),\\\label{vec9}
\mathcal{V}_9\equiv(0,0,\delta_2^2,\gamma_2^2).
\end{align}
\end{subequations}
Among the above nine vectors if it turns out that four are linearly independent then those four vectors span the whole 2-qubit tensor product Hilbert space and hence there will be no vector to exhibit the Hardy's argument (\ref{Hardy3}). However, from the above expressions it is clear that the set of vectors $\{\mathcal{V}_4,\mathcal{V}_5,\mathcal{V}_6\}$ is linearly independent. If we consider the set $\{\mathcal{V}_4,\mathcal{V}_5,\mathcal{V}_6,\mathcal{V}_7\}$ then it will be linearly dependent provided $\mbox{Det}[\mathcal{V}_4,\mathcal{V}_5,\mathcal{V}_6,\mathcal{V}_7]=-\alpha_1^2\beta_1^2\gamma_2^1=0$. Eq.(\ref{A1}) tells that neither $\alpha_1^2$ nor $\beta_1^2$ can be zero. But, no such restriction applies for $\gamma_2^1$ to be nonzero. In the similar way, analyzing the criteria for linear dependence of the different sets $\{\mathcal{V}_4,\mathcal{V}_5,\mathcal{V}_6,\mathcal{V}_k\}$ with $k=1,2,3,8,9$, we obtain the following conditions:
\begin{equation}
\gamma_1^2=\gamma_2^2=(\alpha_1^2\beta_2^0-\alpha_2^0\beta_1^2)=(\alpha_1^2\beta_2^1-\alpha_2^1\beta_1^2)=0
\end{equation}
For exhibiting the Hardy's argument (\ref{Hardy3}) the concerned 2-qubit state must be orthogonal to the subspace spanned by the set $\{\mathcal{V}_4,\mathcal{V}_5,\mathcal{V}_6\}$ and this unique state reads as:
\begin{equation}
|\psi\rangle_{2\times2}^{Hardy}\propto(\beta_1^2|\psi_1^0\rangle-\alpha_1^2|\psi_1^1\rangle)\otimes|\phi_1^1\rangle.
\end{equation}
As the state turns out to be a product state it cannot manifest the Hardy's argument (\ref{Hardy3}).

{\bf Case(ii)}: Here all the POVM elements are in general not rank one operator. To satisfy the conditions (\ref{H31})-(\ref{H33}) the concerned 2-qubit state must be orthogonal to the subspace spanned by the following nine product operators:
\begin{center}
\begin{table}[h]
\begin{tabular}{|c|c|c|c|c|}
		\hline
	$~~~~\mathcal{E}_2^0\otimes\mathcal{F}_1^1~~~~$ &  $~~~~\mathcal{E}_2^0\otimes\mathcal{F}_1^2~~~~$    &   $~~~~\mathcal{E}_2^1\otimes\mathcal{F}_1^2~~~~$ \\ 
		\hline
	$~~~~\mathcal{E}_1^1\otimes\mathcal{F}_1^0~~~~$  & $~~~~\mathcal{E}_1^2\otimes\mathcal{F}_1^0~~~~$    &   $~~~~\mathcal{E}_1^2\otimes\mathcal{F}_1^1~~~~$\\ 
	\hline
	$~~~~\mathcal{E}_1^0\otimes\mathcal{F}_2^1~~~~$  & $~~~~\mathcal{E}_1^0\otimes\mathcal{F}_2^2~~~~$    &   $~~~~\mathcal{E}_1^1\otimes\mathcal{F}_2^2~~~~$\\
	\hline
\end{tabular}
\caption{Any $2$-qubit state that manifest the Hardy's argument (\ref{Hardy3}) must be orthogonal to the support of each of these nine product operators.}\label{Table1}
\end{table}
\end{center}
If $\mathcal{E}_2^0$ is rank two and all others are rank one then it is straightforward to argue that ranges of these operators together span four dimension of the tensor product Hilbert space $\mathbb{C}^2\otimes\mathbb{C}^2$ and thus there is no space left for exhibiting Hardy's argument (\ref{Hardy3}). Similar is true for other cases. In some cases (e.g. $\mathcal{E}_1^0$ is rank two and rest are rank one) ranges of these operators together span three dimension and in such cases the analysis boils down to the Case(i). 

Form the analysis so far presented, it is clear that no 2-qubit state exhibits the three-outcome Hardy's argument. This fact provides information about the Hilbert space dimension of the composite system's, i.e., any system that manifests the Hardy's argument (\ref{Hardy3}) cannot be a $\mathbb{C}^2\otimes\mathbb{C}^2$ system. At this point one cannot make any comment about the Schmidt rank of the system, since a $\mathbb{C}^2\otimes\mathbb{C}^3$ state may also have Schmidt rank $2$ and can exhibit the Hardy paradox (\ref{Hardy3}). Similar analysis shows that no $\mathbb{C}^2\otimes\mathbb{C}^2$ exhibits the Hardy paradox (\ref{Hardy3}) (see the Appendix).

\begin{figure}[t!]
\centering
\includegraphics[height=5cm,width=7cm]{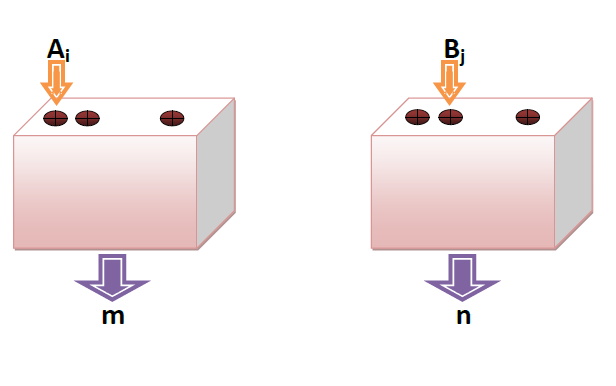}
\caption{Black-box description of the experiment that witness the Schmidt rank in device-independent manner.}\label{fig1}
\end{figure}

Moving further it will be now shown that the Hardy's argument (\ref{Hardy3}) is useful for witnessing the minimal Schmidt rank in device independent manner. In device independent scenario one  does not have detailed knowledge about the experimental apparatus. So a black-box description of the experiment has been shown in Fig(\ref{fig1}). On each side, the experimental device is depicted like a box with some knobs. A knob with different positions on each device, denoted respectively by $A_i$ and $B_j$, allows Alice and Bob to change the parameters of each measuring apparatus.  
Each measurement performed by Alice and Bob has $d$ possible outcomes. Finally, the frequencies $P(A_i=m,B_j=n)$ of occurrence of a given pair of outcomes for each pair of measurements have been collected. After some calculations with the observed frequency the aim is to make some conclusion about the Schmidt rank of the state shared between the two devices.  
 
For a given unknown bipartite state Alice and Bob are asked to perform two different measurements with three outcomes. The resulting statistics $\{P(A_i=m,B_j=n)\}_{i,j=1,2}^{m,n=0,1,2}$ have been collected. It will be checked whether the collected statistics satisfy the conditions described in Eq.(\ref{Hardy3}). According to these conditions the left hand side of (\ref{H34}) must be strictly greater than zero. Among the three terms of (\ref{H34}) if one is nonzero (say, $P(A_2=0,B_2=1)>0$) and the rest two are zero then also the required condition is satisfied. It is important to note that sharing a $2$-qubit pure entangled state \cite{Kar} (i.e. a state with Schmidt rank two) and performing suitable two-outcome measurements on each side the required conditions can be satisfied. Here each of Alice and Bob will assign zero probability for the third outcome. In this way the maximum value of left hand side of (\ref{H34}) can reach up to $0.09$  \cite{Rabelo}. One can impose a more stringent restriction -- that all the three terms in the left hand side of (\ref{H34}) should be nonzero. The previous analysis tells that in quantum theory this stringent conditions cannot be satisfied by performing three-outcome generalized measurements on $2$-qubit (qubit-qutrit) state. However, Alice and Bob can have the following strategy. Suppose they share the following higher dimensional state:
\begin{equation}\label{rho}
\rho=\sum_{k=1}^{3}p_k|{\xi_k}\rangle_{AB}\langle\xi_k|\otimes|kk\rangle_{A'B'}\langle kk|,
\end{equation}
where the particles $A$ and $A'$ are in Alice's lab and the particles $B$ and $B'$ are in Bob's lab. Each $|\xi_k\rangle$ is a $2$-qubit pure entangled states and hence each of them are of Schmidt rank two. Here the primed particles behave as \emph{flag} variable. Whenever the primed particles are in the state $|kk\rangle_{A'B'}$ (which Alice and Bob can know by performing a Von neumann measurement in $\{|k\rangle\}_{k=1}^3$ basis) then Alice and Bob certainly know that the unprimed particles are in the state $|{\xi_k}\rangle$. Note that in the bipartition $AA'$ vs $BB'$ the Schmidt number is two. If the state is $|\xi_1\rangle$, then on their respective particle (unprimed) they perform suitable two-outcome measurements, which exhibits the two-outcome Hardy's argument (\ref{Hardy}) and rename the outcomes accordingly. Thus they are able to make the first term in the left hand side of (\ref{H34}) nonzero. Similarly, when the unprimed particles are in the state $|\xi_2\rangle$ ($|\xi_3\rangle$), Alice and Bob can make the second (third) term in (\ref{H34}) nonzero by performing suitable measurements and renaming the outcomes accordingly \cite{self,cryp}. Sharing this type of states the stringent conditions, that all the terms in (\ref{H34}) are nonzero, can be satisfied. But, due to convexity the success probability cannot be greater than $0.09$. Thus, whenever the success probability is strictly greater than $0.09$, the shared state must have Schmidt rank greater than $2$. This provides the information about the minimal Schmidt rank of the shared bipartite system and importantly it has been done in device independent manner. It also gives information about the minimal dimension of the shared quantum system, that the resulting statistics cannot be obtained from a $2$-qubit or a qubit-qutrit state.  

Besides revealing the the contradiction of quantum mechanics with local-realism Hardy's argument also finds applications in various information theoretic tasks. It has been proved to be useful in witnessing post quantum correlations \cite{Das}. In the recent times various device-independent \cite{Scarani} information theoretic protocols like cryptography \cite{cry1}, randomness certification \cite{rand1}, Hilbert Spaces dimension witness \cite{Brunner} make use of nonlocality arguments \cite{nonlocality}. In this work it has been shown that such an argument turns out to be useful for inspecting the minimal Schmidt rank as well as the minimal Hilbert space dimension in device independent manner.

{\bf Acknowledgments}: It is a pleasure to thank Guruprasad Kar for various simulating discussions and useful suggestions. We also thank Sibasish Ghosh, Samir Kunkri, and Ramij Rahaman for useful discussions. AM acknowledge support from the CSIR project 09/093(0148)/2012-EMR-I.

\section*{Appendix: $\mathbb{C}^2\otimes\mathbb{C}^3$ scenario}
Consider that Alice holds the qubit system and Bob holds the qutrit system. Like in the $2$-qubit scenario here also the following two cases are possible: (A-i) All the POVM elements are rank one operators; (A-ii) Some of the POVM elements may have more than one rank.

{\bf Case(A-i)}: Alice and Bob perform three-outcome rank one POVMs on their respective parts of the shared qubit-qutrit state. Like the $2\times2$ scenario, consider $\{|\psi_1^0\rangle,|\psi_1^1\rangle\}$ as basis for the Alice's qubit system. For Bob's qutrit system consider $\{|\phi_1^0\rangle,|\phi_1^1\rangle,|\phi_1^2\rangle\}$ as basis. Then the other vectors on Bob's side can be expressed as:
\begin{equation}\label{B31}
|\phi_2^n\rangle=\delta_2^n|\phi_1^0\rangle+\gamma_2^n|\phi_1^1\rangle+\eta_2^n|\phi_1^2\rangle. 
\end{equation} 
In this case $\{|\psi_1^0\rangle\otimes|\phi_1^0\rangle,|\psi_1^0\rangle\otimes|\phi_1^1\rangle,|\psi_1^0\rangle\otimes|\phi_1^2\rangle, |\psi_1^1\rangle\otimes|\phi_1^0\rangle,|\psi_1^1\rangle\otimes|\phi_1^1\rangle,|\psi_1^1\rangle\otimes|\phi_1^2\rangle\}$ forms a basis for the $\mathbb{C}^2\otimes\mathbb{C}^3$ tensor product Hilbert space. According to the conditions (\ref{H31})-(\ref{H33}) the qubit-qutrit state exhibiting Hardy's test (\ref{H3}) must be orthogonal to the following nine vectors:
\vspace{-.5cm}
\begin{subequations}
\begin{align}\label{vec1_2*3}
\mathcal{V}_1\equiv(0,\alpha_2^0,0,0,\beta_2^0,0),\\\label{vec2_2*3}
\mathcal{V}_2\equiv(0,0,\alpha_2^0,0,0,\beta_2^0),\\\label{vec3_2*3}
\mathcal{V}_3\equiv(0,0,\alpha_2^1,0,0,\beta_2^1),\\\label{vec4_2*3}
\mathcal{V}_4\equiv(0,0,0,1,0,0),\\\label{vec5_2*3}
\mathcal{V}_5\equiv(\alpha_1^2,0,0,\beta_1^2,0,0),\\\label{vec6_2*3}
\mathcal{V}_6\equiv(0,\alpha_1^2,0,0,\beta_1^2,0),\\\label{vec7_2*3}
\mathcal{V}_7\equiv(\delta_2^1,\gamma_2^1,\eta_2^1,0,0,0),\\\label{vec8_2*3}
\mathcal{V}_8\equiv(\delta_2^2,\gamma_2^2,\eta_2^2,0,0,0),\\\label{vec9_2*3}
\mathcal{V}_9\equiv(0,0,0,\delta_2^2,\gamma_2^2,\eta_2^2).
\end{align}
\end{subequations}
Among the above nine vectors the set $\{\mathcal{V}_1,\mathcal{V}_2,\mathcal{V}_4,\mathcal{V}_5,\mathcal{V}_7\}$ are linearly independent.And the rest four vectors (i.e. $\{\mathcal{V}_k|k=3,6,8,9\}$) can be expressed in terms of these vectors provided the following conditions are satisfied: 
\begin{equation}
\eta_2^1=\eta_2^2=\gamma_2^1=0.
\end{equation}
The unique qubit-qutrit state orthogonal to the subspace spanned by the set $\{\mathcal{V}_1,\mathcal{V}_3,\mathcal{V}_4,\mathcal{V}_5,\mathcal{V}_8\}$ reads as:
\begin{equation}
|\psi\rangle_{2\times3}^{Hardy}\propto(\alpha_1^2|\psi_1^0\rangle-\beta_1^2|\psi_1^1\rangle)\otimes|\phi_1^1\rangle
\end{equation}
Being a product state the above qubit-qutrit state cannot manifest the Hardy's argument (\ref{Hardy3}).

{\bf Case(A-ii)}: In this case some POVM elements are greater than rank one operators. The analysis goes similar as Case(ii). The $\mathbb{C}^2\otimes\mathbb{C}^3$ state exhibiting Hardy's argument (\ref{Hardy3}) need to be orthogonal to the support of the each product operator in Table-(\ref{Table1}). For some cases the ranges of theses operators together span the six dimension of the $\mathbb{C}^2\otimes\mathbb{C}^3$ Hilbert space and hence in such cases, there is no possibility for Hardy's state. For rest of the cases the ranges together span five dimension and hence the cases boil down to the Case(A-i).


\begin{thebibliography}{99}
\bibitem{self} At this point it is important to note that, in the case of BB84 cryptography protocol if the required correlation results from measurements in the $x$ and $z$ bases on qubit pairs then security follows only if it is obtained from $2$-qubit maximally entangled state. However, the same correlation can also be obtained from higher dimensional separable state and the security analysis breaks down (see \cite{cryp}).
\end{thebibliography}
\end{document}